# Can My Microservice Tolerate an Unreliable Database? Resilience Testing with Fault Injection and Visualization

Michael Assad
michael.assad@tum.de
Technical University of Munich
Munich, Germany

Christopher S. Meiklejohn
Heather Miller
cmeiklej@cs.cmu.edu
heather.miller@cs.cmu.edu
Carnegie Mellon University
Pittsburgh, PA, USA

Stephan Krusche
krusche@tum.de
Technical University of Munich
Munich, Germany

## Abstract

In microservice applications, ensuring resilience during database or service disruptions constitutes a significant challenge. While several tools address resilience testing for service failures, there is a notable gap in tools specifically designed for resilience testing of database failures. To bridge this gap, we have developed an extension for fault injection in database clients, which we integrated into Filibuster, an existing tool for fault injection in services within microservice applications. Our tool systematically simulates database disruptions, thereby enabling comprehensive testing and evaluation of application resilience. It is versatile, supporting a range of both SQL and NoSQL database systems, such as Redis, Apache Cassandra, CockroachDB, PostgreSQL, and DynamoDB. A defining feature is its integration during the development phase, complemented by an IntelliJ IDE plugin, which offers developers visual feedback on the types, locations, and impacts of injected faults. A video demonstration of the tool's capabilities is accessible at https://youtu.be/bvaUVCy1m1s.

## 1 Introduction

The software development landscape has witnessed a recent profound shift towards microservice architecture. This decentralized approach, where applications are constructed as a collection of loosely coupled, independent services, has garnered popularity. Microservice applications are typically composed of multiple services and their corresponding databases [9]. The architecture promises scalability, flexibility, and the ability to build and deploy services independently [2, 6].

Yet, they are not without their own set of challenges. As services multiply and inter-service communication grows, the complexity of managing these interactions and ensuring application resilience escalates [11]. The decentralized nature of microservices introduces new failure modes, where only a subset of the services or databases in a microservice application fail.

So, how can developers ensure these service failures are handled gracefully and do not lead to application disruptions?

The emerging answer in the industry is fault injection using chaos engineering [1]. This approach involves intentionally injecting faults into live systems to test and improve their resilience against failures.

While there are several tools available for fault injection, including ChaosMesh[1], Gremlin's[2] ALFI[3][3], Litmus[4], RainMaker[4], and Filibuster[5], their primary focus is on introducing faults into services. However, underexplored remains fault injection that targets the database clients within the microservice applications. While it is crucial to test how the application reacts to service failures, it is equally important to understand the impacts when the databases these services rely on experience issues.

Most fault injection tools approach databases as if they were just another set of containerized services. This perspective allows them to inject standard faults like terminating the container instance, introducing network latency, or inducing CPU overuse in certain nodes. However, this approach has a limitation: it cannot inject faults specific to database clients. It merely focuses on the general container layer. Moreover, in many real-world scenarios, databases do not run within containers. Consequently, these tools, by only targeting containerized environments, are incapable of effectively testing the resilience of systems where databases exist outside of containers.

Developers have limited options when it comes to testing the resilience of their applications with databases running outside of containers. AWS Aurora offers fault injection into database queries[6].

---

[1]ChaosMesh. https://chaos-mesh.org/. Last accessed: October 23, 2023.
[2]Gremlin. https://www.gremlin.com/. Last accessed: October 23, 2023
[3]ALFI was abandoned as of February 2022.
[4]Litmus. https://litmuschaos.io/. Last accessed: October 23, 2023.
[5]Filibuster. https://www.filibuster.cloud/. Last accessed: October 23, 2023.
[6]AWS Aurora. https://docs.aws.amazon.com/AmazonRDS/latest/AuroraUserGuide/AuroraMySQL.Managing.FaultInjectionQueries.html. Last accessed: October 23, 2023.

It can simulate replication and disk failures, as well as force crashes of the database instance. Testing with Aurora requires hosting the database on AWS. Forcing a crash of the database instance could have unintended consequences such as data corruption, prolonged unavailability, or, in worst-case scenarios, irreversible data loss.

In this paper, we introduce a tool designed for injecting faults into the database clients running within microservice applications, without manipulating the actual database. We built this tool as an extension to FILIBUSTER. Our contribution can simulate genuine exceptions that are specific to the database client. Furthermore, it can emulate situations where databases return corrupted data to queries — *Byzantine faults* [5]. This work aims to fill the existing gap in the domain, offering a more comprehensive solution for resilience testing of microservice applications with database interactions.

## 2 Service-Level Fault Injection Testing

An alternative to *chaos engineering* is *Service-Level Fault Injection Testing*. *SFIT* is a technique that integrates fault injection testing into the early stages of the development process. It is inspired by *LFI* (library fault injector) [7]. Unlike chaos engineering, *SFIT* can inject faults prior to the application deployment [8]. Using this technique, developers can inject faults into the services within their microservice application and test its resilience, directly in their IDE.

*SFIT* allows for fine-grained fault injection that targets individual call sites, without necessitating the shutdown of an actual service. In particular, *SFIT* does not inject the fault into the actual running service. Instead, it intercepts the request before it reaches the service, and directly injects the fault.

Given a passing end-to-end functional test that verifies the system behavior, *SFIT* intercepts the gRPC or HTTP calls between the microservices. Then, it enumerates all the possible errors that can occur. *SFIT* re-executes the functional test while injecting one or more errors in the gRPC/HTTP invocations between the services.

As a result of the injected faults, the functional test might fail. Here would be the developer's opportunity to add *fault injection predicates* to their test. These allow developers to write conditional expressions of varying precision that encode the expected behavior of the application when a fault is present.

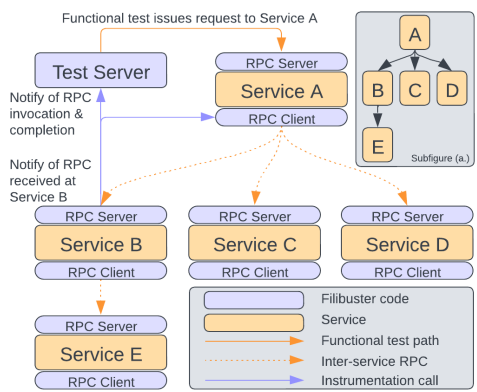

**Figure 1:** *SFIT* **instrumentation of RPCs in a microservice application. (a) The original microservice application.**

FILIBUSTER[7] is the prototype implementation of *SFIT*. Figure 1 shows how FILIBUSTER instruments an example microservice application. Subfigure 1a shows the original microservice application.

Input to FILIBUSTER is an end-to-end functional test that initially invokes service A. Service A further calls B, C, and D. Service B in turn calls E. By using FILIBUSTER, the RPC between services B and E will undergo testing for all potential RPC error status codes that are pre-defined in FILIBUSTER. These faults can be, for example, the gRPC error statuses *DEADLINE_EXCEEDED* or *NOT_FOUND*.

Then, FILIBUSTER injects faults in the RPCs between A and its immediate downstream services B, C, and D. FILIBUSTER proceeds to inject faults covering all possible combinations of failures occurring simultaneously in multiple RPC invocations.

## 3 Fault Injection in Database Clients: Key Features

While FILIBUSTER can provoke error handling code in a service's dependencies to see how a service reacts under downstream service failure, what remains unknown is how services respond to their upstreams when their database queries throw or return invalid values.

With nearly half of the microservice applications relying on databases [10], there is a growing need for tools that can inject faults specific to the database client. Such an extension would not only make FILIBUSTER more comprehensive but also equip developers with the means to test the resilience of their applications in scenarios where the database experiences issues, which is our research goal.

Thus we ask: **what kind of faults can databases experience?** Broadly, they can be categorized into two main types. The first pertains to exceptions thrown by the database client. These could arise due to various reasons, such as connectivity issues, query timeouts, or resource constraints. When a database client throws an exception, the service relying on that database should ideally handle the exception properly, ensuring that the end-user experience is not adversely impacted.

The second type is more challenging to detect: data corruption. We refer to this type of fault as *Byzantine fault*. In this case, the database client does not throw any explicit errors. Instead, it returns corrupted data. Data corruption could have many reasons. For instance, it could be due to inconsistency errors in a database cluster, or due to badly managed deployments. Handling such faults requires more careful mechanisms, as the corrupted data can potentially cascade through the system, leading to unpredictable and often erroneous behaviors.

We present a tool for fault injection in database clients within microservice applications. Faults are injected during the development phase, directly at the IDE. We built this tool as an extension to FILIBUSTER. While also extendable to non-microservice applications, the true efficacy of this extension manifests when applied to microservice architectures. Its value becomes apparent in the analysis of complex application behavior when a permutation of services and databases fails. This is particularly relevant in the context of Byzantine faults, wherein a faulty value injected into a database client directly called by a service may propagate and cause application failure downstream in subsequent services.

---
[7]Java open source implementation of FILIBUSTER. https://github.com/filibuster-testing/filibuster-java-instrumentation. Last accessed: October 23, 2023.

Our objective does not involve the identification of bugs within the database or its client per se; rather, we focus on discovering bugs residing in applications utilizing databases. These bugs manifest within the application code. We discover these bugs by deliberately injecting faults into the database client, thereby instigating and identifying latent bugs within the applications interfacing with said database client.

The main features of our extension are:

### 3.1 Instrumentation for Diverse Databases

We provide support to a diverse set of both SQL and NoSQL database systems. Specifically, we can introduce faults in the in-memory Redis[8] data store, Apache Cassandra[9], the relational PostgreSQL[10], the distributed SQL database CockroachDB[11], and locally running Amazon's DynamoDB[12]. This broad compatibility ensures that developers can use FILIBUSTER to test their application's resilience across varied database systems.

### 3.2 Support for Synchronous and Asynchronous APIs

Our fault injection tool can inject faults into both the synchronous and asynchronous APIs of the database clients. In the synchronous API, faults are thrown immediately upon invocation. Conversely, in the asynchronous API, faults manifest later, typically when the returned *future* is resolved. This distinction ensures accurate fault simulation, aligning with the real-world behavior of database interactions.

### 3.3 Customizable Byzantine Faults

Byzantine faults simulate scenarios where the database returns corrupted data. For cache systems, a typical Byzantine fault could involve returning a cache miss, such as null, an empty string, or an empty array, when a cache hit is expected. This could be the case if the cache fails and is restarted for any reason.

Unlike typical database research that often focuses on simulating malfunctions within the database itself, Byzantine fault injection, in this context, seeks to emulate the unintended side effects arising from malfunctioning applications that read and write data. For instance, a service reading from a database might encounter data written by a different service version, resulting in a format discrepancy indistinguishable from data corruption.

Developers can define their own *transformation* functions. These are functions that are applied to the values returned from the database to simulate cases where the data is corrupt. FILIBUSTER comes pre-equipped with several transformation functions. For instance, mutating chars in strings, negating boolean values, or flipping bits in byte arrays. In case the database returns a JSON object, FILIBUSTER can apply transformation functions according to the identified type of each field in the JSON Object. In instances where a JSON object is nested, FILIBUSTER applies the transformation function to all fields, including those within the nested structures.

[8]Redis. https://redis.com/. Last accessed: October 23, 2023.
[9]Apache Cassandra. https://cassandra.apache.org/. Last accessed: October 23, 2023.
[10]PostgreSQL. https://www.postgresql.org/. Last accessed: October 23, 2023.
[11]CockroachDB. https://www.cockroachlabs.com/. Last accessed: October 23, 2023.
[12]DynamoDB. https://aws.amazon.com/de/dynamodb/. Last accessed: October 23, 2023.

### 3.4 Injecting Exceptions

The exceptions injected by FILIBUSTER are not arbitrary. They are crafted to mimic the real exceptions as defined in the documentation of the supported databases. The tool only introduces exceptions at API methods where such exceptions are actually throwable. Additionally, the same exception messages as those detailed in the database documentation are used for injection, ensuring that faults are both realistic and consistent with actual operational challenges. This could be, for example, *timeout* or *unavailable* exceptions. This approach enhances the reliability and relevance of the testing process.

## 4 Integrated IDE Plugin: A Visual Approach

Our tool for database fault injection integrates with FILIBUSTER's IDE plugin. This integration serves as a bridge, enabling developers to visualize both the injected faults and the specific call sites they target.

One of the features of this integrated plugin is its use of color coding. This visual distinction allows developers to quickly differentiate between the two fault injection types: exceptions and Byzantine faults. By glancing at the color cues, developers can identify the type of fault without diving deep into logs or text descriptions.

The decision to embed fault injection visualization directly within the IDE carries significant benefits. Central to these is the immediacy and clarity it provides to the development process. Developers can, as soon as they have executed the FILIBUSTER tests, ascertain all the details about the fault injection. They can determine the specific fault introduced, the call site where it was injected, and, importantly, whether that injected fault led to test failures. Anecdotally, we observed that adoption of FILIBUSTER increased when the first UI was released.

Consider the test below for a basic feature in a social media platform. Upon user login, the system fetches their profile based on a unique username (the provided key), retrieving a JSON object. This JSON object contains a *boolean* field denoting whether the user's account is verified. Additionally, there is a *string* field representing the *username* of the most recently viewed profile. For convenience, the system retrieves the JSON object of the last viewed profile and displays it post-login, allowing users to pick up their browsing from where they left off.

```
@TestWithFilibuster
public void test() {
    JSONObject user = redisSyncCommands.get("john_doe");
    RedisFuture<String> f_last_visited = redisAsyncCommands
            .get(user.getString("last_visited_profile"));
    String last_visited = f_last_visited.get();
    assertEquals("joe_bloggs", last_visited);
}
```

We run that functional test with FILIBUSTER. Figure 2 shows a screenshot of the FILIBUSTER plugin, running in IntelliJ IDEA. In this iteration, two faults are simultaneously injected: a Byzantine fault, and an exception.

(1) The first row in orange indicates the injection of a Byzantine fault. The fault is injected in the RPC method with the fully qualified name `RedisStringCommands/get` that was invoked with the string `john_doe` as an argument, representing the username of the user who is logging in. This corresponds to line three in the functional test. The fault is injected in the RPC response. Filibuster flips the value of the boolean *is_verified* from *true* to *false*. The last field — *Fault Injected?* — shows further information about the injected fault. We show below the content of the *Fault Injected?* cell.

| RPC Method | RPC Arguments | RPC Response | Fault Injected? |
|---|---|---|---|
| Test Block: da39a3ee5e6b4b0d3255bfef95601890afd80709 | | | |
| `RedisStringCommands/get` | `Object;`<br>`[john_doe]` | `Object`<br>`{"last_visited_profile":"joe_bloggs","is_verified":"false","email":"john@gmail.com"}` | `Object`<br>`accumulator = {"referenceValue":"{\"last_visited_profile\":\"joe_bloggs\",\"is_verified\":true,\"email\":\"john@gmail.com\"}","context":[{"key":"last_visited_profile","value":"` |
| `RedisStringAsyncCommands/get` | `Object;`<br>`[joe_bloggs]` | `RedisFuture`<br>`AsyncCommand [type=GET, output=ValueOutput [output=null, error='null'], commandType=io.lettuce.core.protocol.Command]` | `RedisCommandTimeoutException`<br>`code =`<br>`cause = Command timed out after 100 millisecond(s)`<br>`cause_message = undefined`<br>`description = undefined` |
| `Future/get` | `Object;`<br>`[]` | `RedisCommandTimeoutException`<br>`code = undefined` | |

Figure 2: Screenshot of Filibuster plugin running in IntelliJ IDEA. Two faults are injected: a Byzantine fault and an exception.

```
{
  "referenceValue": {
    "last_visited_profile": "joe_bloggs",
    "is_verified": true,
    "email": "john@gmail.com"
  },
  "context": [ {
    "key": "last_visited_profile",
    "value": {
      "referenceValue": "joe_bloggs",
      "context": 9
    } }, {
    "key": "is_verified",
    "value": {
      "referenceValue": true
    } } ]
}
```

The original values fetched from the database are referred to as the *reference values*. Since the database originally returns a JSON object with multiple fields, Filibuster attached different transformers to each field, based on its identified data type. Specifically, for the *string* field *last_visited_profile* with the reference value *joe_bloggs*, Filibuster attaches a transformer that iteratively mutates its characters. Here, *context* refers to the index of the mutated char. The *context* field carries pertinent information about current and previously injected Byzantine faults. All mutations of the *chars* in that *string* were exhausted in previous Filibuster iterations. In the present iteration, Filibuster has flipped the *boolean* value associated with the *is_verified* field. Notably, this particular transformation does not necessitate any supplementary *context*, as the reference value is sufficient to indicate what the transformation should be.

(2) The second row in red shows the call site where Filibuster injected a RedisCommandTimeoutException. The exception is injected in the method RedisStringAsyncCommands/get. This corresponds to lines four and five in the functional test. However, that method does not throw immediately, since it is asynchronous. It throws when its associated future is resolved. The third row in light yellow shows the call site where the *future* was resolved, and the fault was injected. This corresponds to line six in the functional test.

**Dynamic Proxy Interceptor.** Central to instrumenting database calls and injecting faults in them is the development of an interceptor that can interpose on call sites and manipulate their response. The interceptor we built utilizes Java's dynamic proxy API, allowing it to be used generically with all the database clients. This ensures that, in the future, the same interceptor can be used to instrument other database clients that we currently do not support. Filibuster does not inject faults into the actual database. Instead, the *dynamic proxy interceptor* intercepts the incoming request before it even reaches the database client, and directly returns a response with the injected fault. To inject faults, developers need to exchange the used database client in their application with a Filibuster instrumented version. In case a dependency injection framework is used, this could amount to only a few changes in the configuration file. The public Filibuster class DynamicProxyInterceptor offers the static method `<T> T createInterceptor(T target, String connectionString)` which returns an intercepted version of a target interface of type T.

## 5 Conclusion

We used Filibuster to inject faults in the database clients of the microservice application of one of the top food delivery services in the USA. Our experiments successfully introduced faults and, notably, replicated database failure conditions that had historically caused application outages. This emphasizes the significance of fault injection as a preventive measure in system resilience. With proactive fault injection testing, several real-world outages could potentially have been circumvented. In a previous evaluation of Filibuster on a corpus of four industrial microservice applications, Filibuster was found to reduce the time needed for manual test creation [8]. It also provided higher code coverage by automatically mocking complex scenarios where multiple services simultaneously fail.

Partial system failures in microservice applications are challenging to understand and debug. When a subset of the services or databases encounter disruptions, the resulting behaviors of the system can be hard to predict. Addressing this, we introduce an extension to Filibuster that systematically injects faults into database clients within microservice applications. Through fault injection, our work simulates realistic failure scenarios, allowing for the evaluation of the application's behavior under disruptions. Developers can view the results of the fault injection in an integrated IDE plugin, thus enhancing the usability and the user experience.